\documentclass[twocolumn,aps,prl,a4paper,preprintnumbers,showpacs,showkeys]{revtex4-1}

\usepackage{amssymb}
\usepackage{amsmath}
\usepackage{graphicx}
\usepackage{dcolumn}
\usepackage{bm}
\usepackage{epsfig}
\usepackage{hyperref}
\usepackage[caption=false]{subfig}    

\newcommand\beq{\begin{eqnarray}}
\newcommand\eeq{\end{eqnarray}}

\newcommand\figwidth{.43\textwidth}
\newcommand\Eq[1]{Eq.~\ref{eq:#1}}
\newcommand\Fig[1]{Fig.~\ref{fig:#1}}

\newcommand\Tab[1]{Table~\ref{tab:#1}}

\newcommand\bfn{\mathbf n}

\newcommand\bfe{\mathbf e}

\newcommand\calL{\mathcal L}
\newcommand\calO{\mathcal O}

\begin{document}

\preprint{RIKEN-QHP-49}

\title{Transdimensional equivalence of universal constants for Fermi gases at unitarity}

\author{Michael G. Endres}
\email{endres@riken.jp}

\affiliation{Theoretical Research Division, RIKEN Nishina Center, Wako, Saitama 351-0198, Japan}


\date{\today}
 
\begin{abstract}
I present lattice Monte Carlo calculations for a universal four-component Fermi gas confined to a finite box and to a harmonic trap in one spatial dimension.
I obtain the values $\xi_{1d} = 0.370(4)$ and $\xi_{1d} = 0.372(1)$, respectively, for the Bertsch parameter, a nonperturbative universal constant defined as the (square of the) energy of the untrapped (trapped) system measured in units of the free gas energy.
The Bertsch parameter obtained for the one-dimensional system is consistent to within $\sim1$\% uncertainties with the most recent numerical and experimental estimates of the analogous Bertsch parameter for a three-dimensional spin-1/2 Fermi gas at unitarity.
The finding suggests the intriguing possibility that there exists a universality between two conformal theories in different dimensions.
To lend support to this study, I also compute ground state energies for four and five fermions confined to a harmonic trap and demonstrate the restoration of a virial theorem in the continuum limit.
The continuum few-body energies obtained are consistent with exact analytical calculations to within $\sim1.0$\% and $\sim0.3$\% statistical uncertainties, respectively.
\end{abstract}

\maketitle

Universal Fermi gases have been the subject of intense study in recent years.
Perhaps the most interesting example of such a system in three spatial dimensions is the unitary Fermi gas: a dilute mixture of spin-1/2 fermions with an attractive short-range two-body interaction tuned to infinite scattering length.
Unitary fermions have been realized in ultra-cold atom experiments by exploiting properties of Feshbach resonances, where the scattering length may be tuned arbitrarily with the application of an external magnetic field \cite{O'Hara13122002,PhysRevLett.93.050401,2003Natur.424...47R,PhysRevLett.91.080406,PhysRevLett.89.203201} (also see \cite{RevModPhys.80.1215}).
In the unitary limit, the details of the interparticle potential become irrelevant, and consequently physical quantities may be characterized by a single dimensionful parameter, the density of the system.

From a theoretical standpoint, the unitary Fermi gas is described by a nonrelativistic conformal field theory \cite{PhysRevD.76.086004}, and despite its simplicity, exhibits a rich and fascinating variety of physical phenomena.
By dimensional analysis, the energy of the untrapped gas at density $\rho$ is given by
\begin{eqnarray}
E(\rho) = \xi_{3d} E_0(\rho)\ ,
\label{eq:untrapped_energy_relation}
\end{eqnarray}
where $\xi_{3d}$ is a nonperturbative universal constant known as the Bertsch parameter \cite{baker2000mbx} and $E_0(\rho)$ is the energy of the noninteracting Fermi gas evaluated at the same density.
Density-functional theory predicts that when confined to a harmonic trap, the energy of the system is given by
\begin{eqnarray}
E^{osc}(Q) = \sqrt{ \xi_{3d} } E^{osc}_0(Q)\ ,
\label{eq:trapped_energy_relation}
\end{eqnarray}
where $E^{osc}_0(Q)$ is the energy of the noninteracting Fermi gas confined to a trap and $Q$ is the total number of fermions, taken to be asymptotically large \cite{PhysRevA.72.041603}.
This result, including subleading corrections in $1/Q$, has been been reaffirmed using a general coordinate-invariant effective field theory description of the system \cite{Son:2005rv}.

In recent years, considerable effort has been devoted to determining the Bertsch parameter to high precision \footnote{See, e.g., \cite{Endres:2012cw} for a extensive summary of experimental, theoretical and numerical estimates for $\xi_{3d}$.}, motivated in part by the fact that a theoretical determination of $\xi_{3d}$ may be directly compared with experimental data (or vice verse).
The current best determination of the Bertsch parameter from exact Monte Carlo simulations is $\xi_{3d} = 0.372(5)$ \cite{PhysRevA.84.061602}, and is in good agreement with the current best experimental value of $\xi_{3d} = 0.376(4)$ \cite{Ku03022012}.
Conformal and scale invariance implies an operator-state correspondence which relates the conformal dimensions of primary operators in free space to the spectrum of harmonically trapped fermions \cite{PhysRevD.76.086004}.
Knowledge of the Bertsch parameter might therefore teach us something about the conformal dimensions of many-body operators in the untrapped theory.

Recently, Nishida and Tan \cite{PhysRevLett.101.170401} observed that a dilute four-component Fermi gas in one spatial dimension is also universal when an attractive four-body interaction is tuned to produce a four-body zero-energy bound state.
The system is conformal, and shares many properties in common with the unitary Fermi gas in three dimensions \cite{Nishida:2009pg}.
Particularly, this system is characterized by a single dimensionful parameter, the density, and as such the untrapped system obeys the same energy relation as \Eq{untrapped_energy_relation}, although with a constant of proportionality $\xi_{1d}$ which need not be the same as $\xi_{3d}$.
A straight-forward calculation using Thomas-Fermi theory predicts that the trapped one-dimensional system will obey \Eq{trapped_energy_relation}, with proportionality constant $\sqrt{\xi_{1d}}$.
In addition to an operator-state correspondence, features shared by both the one- and three-dimensional Fermi gases include Universal (Tan) relations, and a virial theorem for fermions confined to a harmonic trap \cite{2008AnPhy.323.2952T,2008AnPhy.323.2971T,2008AnPhy.323.2987T,PhysRevLett.100.205301,2008PhRvA..78e3606B}.

\begin{figure*} 
\begin{tabular}{cc}
\includegraphics[width=\figwidth]{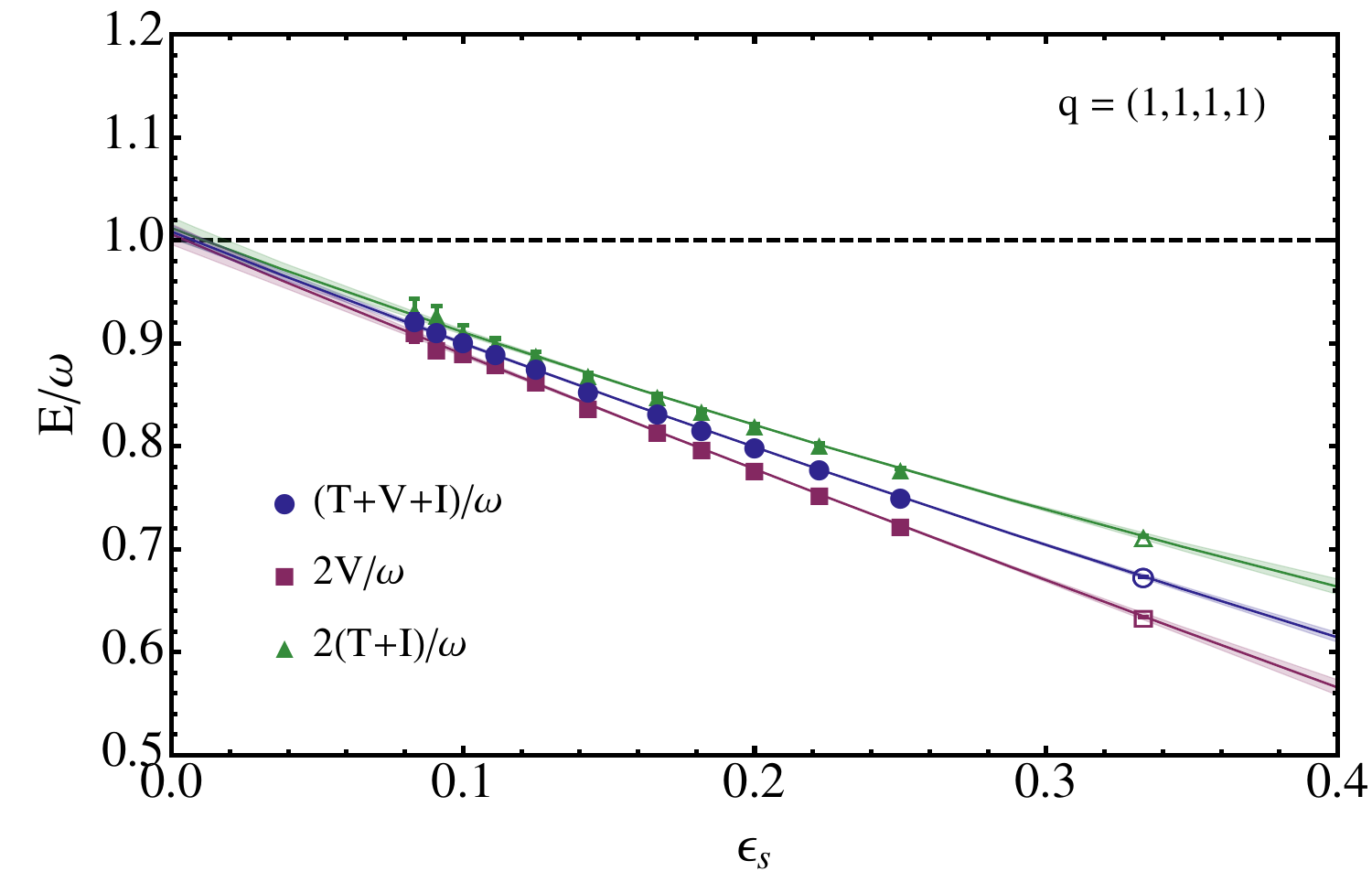}
\includegraphics[width=\figwidth]{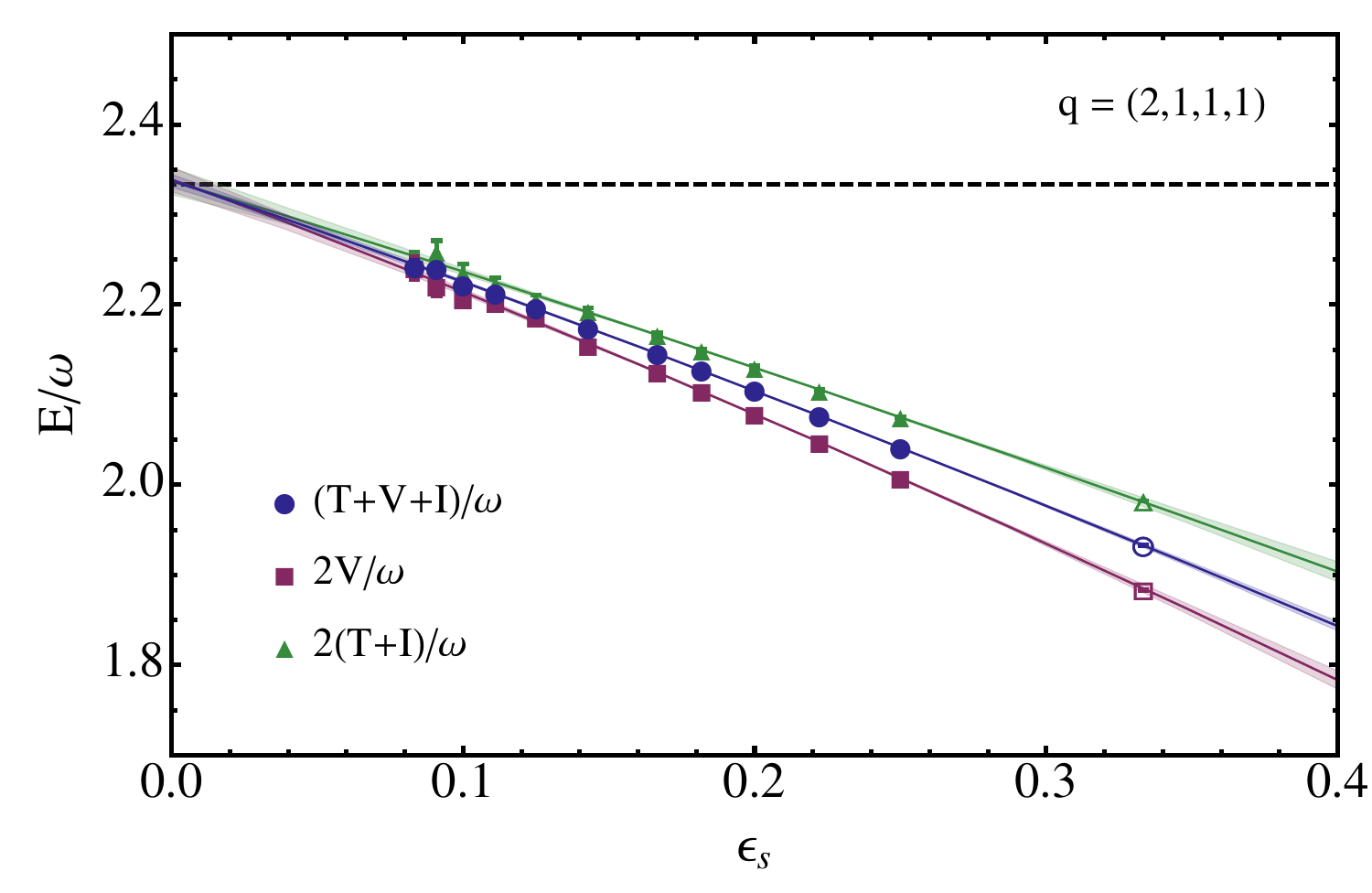}
\end{tabular}%
\caption{\label{fig:few-body}%
Continuum extrapolation (using filled data points only) of energies for four (left) and five (right) fermions confined to a harmonic trap.
Dashed lines indicate the exact values for the energy in the continuum.} 
\end{figure*}

In this work, I present results from lattice Monte Carlo simulations of the one-dimensional Fermi gas confined to a finite box and a harmonic trap.
I verify that trapped few-body systems comprising four and five fermions have ground state energies consistent with exact theoretical predictions, and also that these states obey a Viral theorem in the continuum limit.
I then present numerical results for many-body systems, determining the value of Bertsch parameter $\xi_{1d}$ in the infinite volume and thermodynamic limits.
Although there is no known theoretical argument to suggest that $\xi_{1d}$ and $\xi_{3d}$ are the same, I show evidence that these two quantities are in fact equal to within 1\% uncertainties.
Finally, I discuss the implications of this finding.


The starting point for this study is a continuum effective field theory for nonrelativistic fermions in two-dimensional Euclidean space-time, described by the Lagrangian \cite{Nishida:2009pg}:
\begin{eqnarray}
\calL = \psi^\dagger \left( \partial_\tau - \frac{\nabla^2}{2m} - \mu \right) \psi - \frac{g}{4!} (\psi^\dagger \psi)^4 \ ,
\label{eq:continuum_action}
\end{eqnarray}
where $\psi_\sigma$ is a four-component Grassmann-valued spinor with components labeled by $\sigma$, $m$ is the fermion mass, $\mu_\sigma$ is a chemical potential coupled to the fermion number for each component, and $g$ is an attractive coupling associated with a four-body contact interaction.
The continuum theory may be discretized on an $N_\tau \times (2N_s+1)$ lattice with sites labeled by the integer coordinate pair $\bfn=(n_\tau,n_s)$, where $n_\tau \in [ 0, N_\tau)$ and $n_s\in [ -N_s, N_s]$ following the approach of \cite{PhysRevLett.92.257002}.
The lattice discretization is carried out by identifying the various terms in \Eq{continuum_action} with their discrete counterparts:
\begin{eqnarray}
\partial_\tau \psi - \mu\psi & \to  & \frac{1}{b_\tau} \left( \psi_\bfn - e^{b_\tau \mu} \psi_{\bfn-{\bfe_\tau}} \right) \ ,\cr
-\nabla^2 \psi               & \to  & \frac{1}{b_s^2} \left( 2\psi_\bfn -\psi_{\bfn+{\bfe_s}} - \psi_{\bfn-{\bfe_s}} \right) \ ,\cr
\psi^\dagger \psi            & \to  & \psi_\bfn^\dagger e^{b_\tau \mu} \psi_{\bfn-{ \bfe_\tau}} \ ,
\end{eqnarray}
where $b_\tau$ and $b_s$ are the temporal and spatial lattice spacings, and $\bfe_\tau$ and $\bfe_s$ are corresponding unit basis vectors.
I impose antiperiodic boundary conditions in the time direction with temporal extent (i.e., inverse temperature) $\beta = b_\tau N_\tau$, and open boundary condition is the space direction with spatial extent $L = b_s (2 N_s + 1)$.
Throughout I work in units where $\hbar=1$.

An arbitrary external potential may be introduced in the lattice construction by promoting the chemical potential to one that is space-dependent: $\mu \to \mu + v_\bfn$.
In this study, I consider potentials of the form $v_\bfn = 0$ (i.e., untrapped) and $v_\bfn = \kappa (b_s n_s)^2/2$ (i.e., trapped), where $\kappa$ is a spring constant controlling the strength of the harmonic potential.
In the presence of a harmonic potential and at infinite volume, the only relevant length-scale describing the physical size of the system is the characteristic trap size $L_0 = 1/(m\kappa)^{1/4}$.
To unify the discussion for untrapped and trapped fermions, I define the characteristic  length scale $L_0 \equiv 4L/\pi$ in the former case.

Unitarity is achieved by tuning the coupling $g$ to some critical value $g_c$, corresponding to a zero-energy four-body bound state at infinite volume (with $\kappa=0$).
The coupling may be determined by exact diagonalization of the lattice Hamiltonian, and yields the integral equation:
\begin{eqnarray}
\frac{1}{2\pi \hat g_c} =  \int_{-\pi}^{\pi} \left( \prod_\sigma{\frac{d p_\sigma}{2\pi}} \right)  \frac{\delta(\sum_\sigma p_\sigma) }{\prod_\sigma ( 1 + \Delta_{p_\sigma}/\hat m) -1}\ ,
\label{eq:four_body}
\end{eqnarray}
where $\Delta_p = 2 \sin^2(p /2)$, $\hat g_c = b_\tau g_c / b_s^3$,  and $\hat m = m b_s^2/b_\tau$.
In this limit, the only dimensionful parameters in the theory are the lattice spacings $b_\tau$ and $b_s$, system size $L_0$ and the fermion mass $m$.
All energy scales must therefore be proportional to $\omega = 1/(mL_0^2)$ and all discretization errors must be a function of the dimensionless ratios $\epsilon_s = b_s/L_0$ and $\epsilon_\tau = b_\tau \omega \sim \epsilon_s^2$ for fixed $\hat m$.

Discretization errors in observables may be understood from the viewpoint of a Symanzik action \cite{Symanzik1983187,Symanzik1983205}: a continuum action in which lattice artifacts are quantified by couplings associated with the untuned irrelevant operators that are allowed by the symmetries of the underlying lattice theory.
Volume scaling may be inferred from the scaling dimensions of those untuned operators;
following \cite{Endres:2012cw}, one expects operators of scaling dimension $\Delta$ to induce volume dependence in dimensionless observables $\calO$ that scale as $L_0^{3-\Delta}$ in one spatial dimension.
The leading scaling is expected to behave as \footnote{Note that $\calO_j$ are implicit functions of the fixed dimensionless quantity $\hat m $.}:
\begin{eqnarray}
\calO(\epsilon_s) = \calO_{cont} + \calO_1 \epsilon_s + \calO_{1.666} \epsilon_s^{1.666} + \ldots \ ,
\label{eq:fit_func}
\end{eqnarray}
where the lowest order correction to the continuum result, $\calO_{cont}$, arises from an untuned dimension two ``effective range'' term in the Symanzik action, and the subleading contribution from an odd-parity five-fermion operator with a scaling dimension $\Delta/2 = 2.333$ \cite{Nishida:2009pg}.

For this study, I consider zero-temperature physics in the canonical ensemble, with fermion number $q_\sigma$ for each species and total fermion number $Q = \sum_\sigma q_\sigma$.
From the fugacity expansion, one may show that the canonical ensemble partition function $\hat Z(q) \equiv  e^{-\beta F(q)}$ is related to the grand-canonical partition function via a Fourier transform with respect to an imaginary chemical potential.
At the critical coupling and at sufficiently low temperature, one may define the total energy of the system as the total derivative:
\begin{eqnarray}
E(q) = \lim_{\beta\to\infty} \left. \frac{d F(q) }{d \log \omega } \right|_\textrm{fixed physics} \ ,
\label{eq:energy}
\end{eqnarray}
where all physical length scales are held fixed \footnote{Here, I assume the canonical partition function has been normalized such that $F(0) = 0$}.
It follows that the total energy for the ground-state of the system may be expressed as the sum of three contributions, $E = T+V+I$, where
\begin{eqnarray}
T(q) = \lim_{\beta\to\infty} \frac{ \partial F(q)}{ \partial\log 1/m } \ ,
\label{eq:kinetic}
\end{eqnarray}
\begin{eqnarray}
V(q) = \lim_{\beta\to\infty} \frac{ \partial F(q)}{ \partial\log \kappa } \ ,
\label{eq:potential}
\end{eqnarray}
and
\begin{eqnarray}
I(q) = \left(- \frac{\partial \log g_c}{\partial \log m} \right) \times  \lim_{\beta\to\infty} \frac{ \partial F(q)}{ \partial\log g_c }
\label{eq:interaction}
\end{eqnarray}
are the expectation values of the kinetic, potential and interaction energy operators, respectively.
In the continuum limit, it was shown that the trapped universal Fermi gas obeys the virial theorem: $E=2V$ \cite{Nishida:2009pg}; one therefore has two independent measures of the energy of the system, namely $E = 2V$ and $E=2(T+I)$, the average of which yields \Eq{energy}.



Numerical simulations of the few- and many-body Fermi gas where performed using a fermion world-line representation of the canonical partition function derived in \cite{2012PhRvA..85f3624E}.
In this representation, $\hat Z(q)$ is expressed as a path-integral over all possible non-intersecting, self-avoiding, closed fermion loops, with the fermion number for each species fixed by a constraint on the winding numbers in the time-like direction.
Explicit expressions for the action and observables are provided in \cite{2012PhRvA..85f3624E}.
Metropolis Monte Carlo updating of the configuration space was performed using a local loop-updating scheme which preserves the constraints on the fermion paths.
Simulations were performed using a fixed mass of $ \hat m =1.3$ corresponding to a critical coupling $\hat g_c \approx 3.7237$, and multiple $\epsilon_s$ and $Q$ values to enable continuum and thermodynamic limit extrapolations.
For each ensemble, approximately 150-350 uncorrelated configurations were generated after thermalization.

From properties of the Schroedinger algebra one may show that the spectrum of few- and many-body systems confined to a harmonic trap consist of a tower of levels separated by $2\omega$ \cite{PhysRevD.76.086004}.
Trapped simulations were therefore performed at a temperature $\beta \omega > 10$ to ensure adequate suppression of excited state contamination.
In the case of untrapped many-body studies, the energy splittings are expected to be of order the Fermi energy, $E_F = Q^2 \omega/2$, and therefore for each $\omega$ the temperature was chosen such that $\beta E_F > 10$.

For four and five fermions confined to a trap, estimates of the energies in units of $\omega$ are plotted in \Fig{few-body} as a function of the discretization error $\epsilon_s$.
Continuum extrapolated energies obtained from a three-parameter least-squares fit to \Eq{fit_func} are provided in \Tab{few_body} and are consistent with exactly determined values of unity and $2.333$ to within 1.0\% and 0.3\% uncertainties, respectively. 
All fits yielded a $\chi^2$ per degree of freedom ($d.o.f.$) of $\lesssim 1$.
The exact continuum energies were determined via the operator-state correspondence using scaling dimensions for few-body operators calculated in \cite{Nishida:2009pg}.
Convergence of the energies defined by $2V$ and $2(T+I)$ in the continuum limit demonstrates a restoration of the virial theorem.

\begin{table}[t]
\caption{\label{tab:few_body}%
Exact and extrapolated continuum few-body energies for fermions confined to a harmonic trap.}
\begin{ruledtabular}
\begin{tabular}{ccccc}
Q & $E/\omega$ (exact) & $E/\omega$ & $2V/\omega$ & $2(T+I)/\omega$  \\
\hline
4 & 1 & 1.008(6) & 1.005(10) & 1.007(11) \\
5 & 2.333 & 2.339(7) & 2.331(12) & 2.346(13)\\
 \end{tabular}
\end{ruledtabular}
\end{table}

\begin{figure} 
\includegraphics[width=\figwidth]{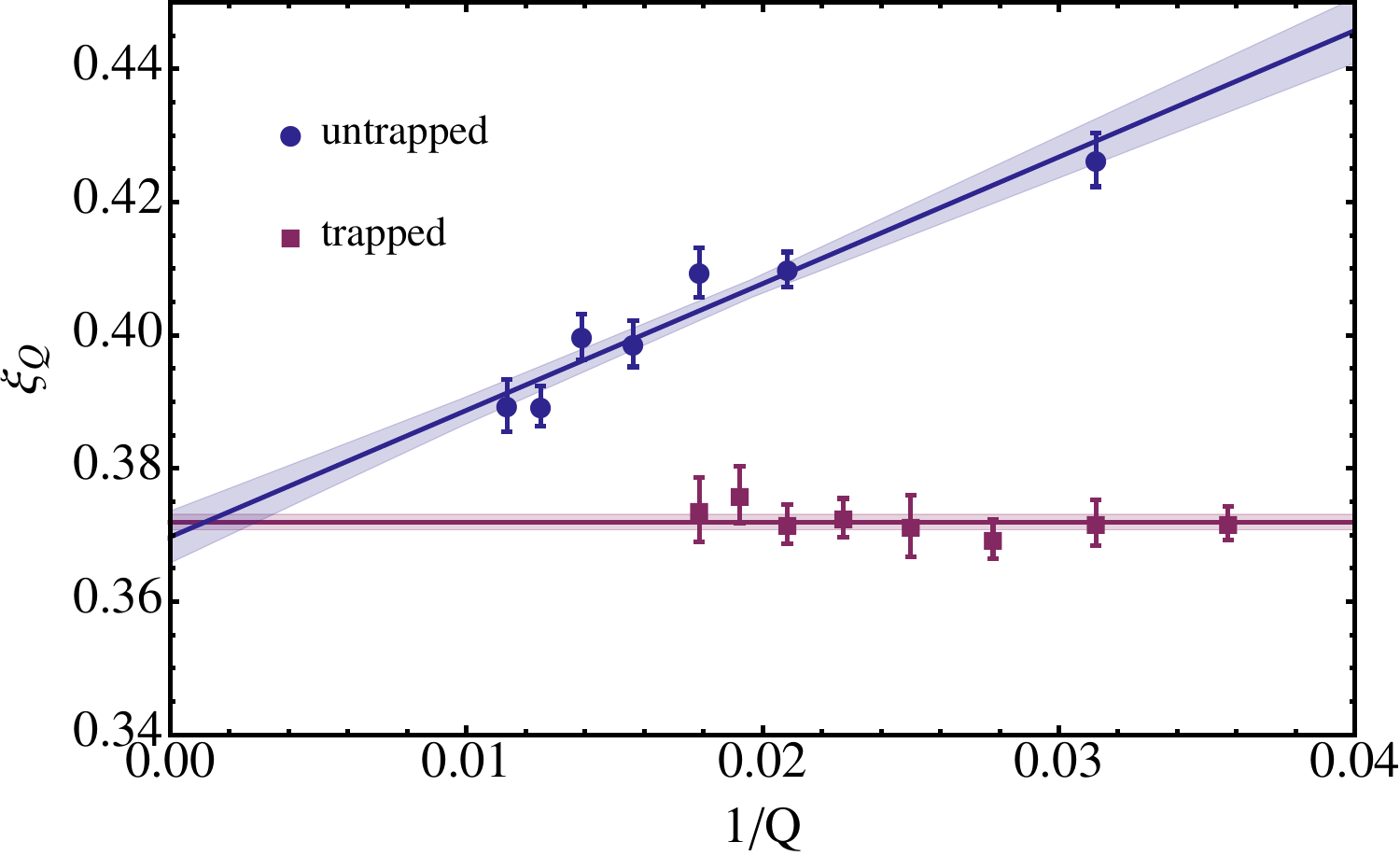}
\caption{\label{fig:bertsch}%
Thermodynamic limit extrapolation of the Bertsch parameter for untrapped and trapped fermions.}
\end{figure}

The one-dimensional Bertsch parameter is defined as the ordered limit $\xi_{1d} = \lim_{Q\to\infty} \lim_{\epsilon_s\to0} \xi_Q(\epsilon_s)$, where $\xi_Q(\epsilon_s)$ is the (square of the) energy of the untrapped (trapped) $Q$-body system at finite $\epsilon_s$ measured in units of the free-gas energy, $E_0 =  Q^3 \omega/24$ ($E_0^{osc} = Q^2\omega /8$).
Untrapped simulations were performed at twelve equally spaced values of $k_F b_s = Q \epsilon_s \in[0.15,0.7]$, where $k_F$ is the Fermi momentum, and for fermion numbers $Q = 32, 48, 56, 64, 72, 80, 88$.
Trapped simulations were performed for all integer values of $1/\epsilon_s \in[7,20]$ and total fermion number $Q = 28, 32, 36, 40, 44, 48, 52, 56$ subject to the constraint $Q^{1/2} \epsilon_s \lesssim 1.0$.
Continuum extrapolations of the energy defined by \Eq{energy} were performed at fixed $Q$ over those respective intervals using \Eq{fit_func} with three fit parameters.
A $\chi^2/d.o.f. \lesssim 1.5$ was achieved over each fit interval and variation of the maximum $\epsilon_s$ value used in the fit yielded statistically consistent extrapolated values for the energies, indicating that the fits were robust.
Thermodynamic limit extrapolations of $\xi_Q(0)$ were subsequently performed.
For the untrapped system, an extrapolation in $1/Q$ was performed using a constant plus linear ansatz fit function, yielding $\xi_{1d}=0.370(4)$ with a $\chi^2/d.o.f. \approx 1.6$.
For the trapped system, an extrapolation was performed using a constant fit function, yielding $\xi_{1d}=0.372(1)$ with a $\chi^2/d.o.f. \approx 0.3$.
Continuum limit extrapolation results for $\xi_Q(\epsilon_s)$ are shown as a function of $1/Q$ in \Fig{bertsch}, along with fit results and error bands for the thermodynamic limit extrapolation of $\xi_Q(0)$.
The values obtained for $\xi_{1d}$ are statistically equal, and moreover, are consistent with the best known Monte Carlo, and experimental values for the three-dimensional Bertsch parameter $\xi_{3d}$.


Provided the interparticle potential is attractive and the system is stable, one may argue that the one- and three-dimensional Bertsch parameters must lie between zero and unity.
The likelihood that the agreement discovered between $\xi_{1d}$ and $\xi_{3d}$ is purely a coincidence is roughly 1/100.
From the point of view of the operator-state correspondence, this equality seems to imply a surprising connection between the conformal dimensions of many-body operators in the one- and three- dimensional universal Fermi gas.
Whether this curious relationship is due to symmetries or is dynamical in origin remains an open question worth pursuing.
If this equality is indeed exact, then numerical simulations of the one-dimensional Fermi gas could offer a computationally inexpensive alternative for computing $\xi_{3d}$ to high precision.
Further exploration of the one-dimensional Fermi gas is underway, including few- and many-body studies of the integrated contact density, and may provide deeper insights into this intriguing relationship.

M.G.E. would like to thank J.-W. Chen, D.B. Kaplan, Y. Nishida, D.T. Son and H. Suzuki for interesting and helpful discussions.
Numerical simulations were conducted on the RIKEN Integrated Cluster of Clusters (RICC) and computer resources provided by the Theoretical High Energy Physics group at Columbia University and the RIKEN BNL Research Center.
M.G.E. is supported by the Foreign Postdoctoral Researcher program at RIKEN and by MEXT Grant-in-Aid for Young Scientists (B) (23740227).

\bibliography{short}

\begin{thebibliography}{27}%
\makeatletter
\providecommand \@ifxundefined [1]{%
 \@ifx{#1\undefined}
}%
\providecommand \@ifnum [1]{%
 \ifnum #1\expandafter \@firstoftwo
 \else \expandafter \@secondoftwo
 \fi
}%
\providecommand \@ifx [1]{%
 \ifx #1\expandafter \@firstoftwo
 \else \expandafter \@secondoftwo
 \fi
}%
\providecommand \natexlab [1]{#1}%
\providecommand \enquote  [1]{``#1''}%
\providecommand \bibnamefont  [1]{#1}%
\providecommand \bibfnamefont [1]{#1}%
\providecommand \citenamefont [1]{#1}%
\providecommand \href@noop [0]{\@secondoftwo}%
\providecommand \href [0]{\begingroup \@sanitize@url \@href}%
\providecommand \@href[1]{\@@startlink{#1}\@@href}%
\providecommand \@@href[1]{\endgroup#1\@@endlink}%
\providecommand \@sanitize@url [0]{\catcode `\\12\catcode `\$12\catcode
  `\&12\catcode `\#12\catcode `\^12\catcode `\_12\catcode `\%12\relax}%
\providecommand \@@startlink[1]{}%
\providecommand \@@endlink[0]{}%
\providecommand \url  [0]{\begingroup\@sanitize@url \@url }%
\providecommand \@url [1]{\endgroup\@href {#1}{\urlprefix }}%
\providecommand \urlprefix  [0]{URL }%
\providecommand \Eprint [0]{\href }%
\@ifxundefined \urlstyle {%
  \providecommand \doi  [0]{\begingroup \@sanitize@url \@doi}%
  \providecommand \@doi [1]{\endgroup \@@startlink {\doibase
  #1}doi:\discretionary {}{}{}#1\@@endlink }%
}{%
  \providecommand \doi  [0]{doi:\discretionary{}{}{}\begingroup
  \urlstyle{rm}\Url }%
}%
\providecommand \doibase [0]{http://dx.doi.org/}%
\providecommand \Doi [0]{\begingroup \@sanitize@url \@Doi }%
\providecommand \@Doi  [1]{\endgroup\@@startlink{\doibase#1}\@@Doi}%
\providecommand \@@Doi [1]{#1\@@endlink}%
\providecommand \selectlanguage [0]{\@gobble}%
\providecommand \bibinfo  [0]{\@secondoftwo}%
\providecommand \bibfield  [0]{\@secondoftwo}%
\providecommand \translation [1]{[#1]}%
\providecommand \BibitemOpen [0]{}%
\providecommand \bibitemStop [0]{}%
\providecommand \bibitemNoStop [0]{.\EOS\space}%
\providecommand \EOS [0]{\spacefactor3000\relax}%
\providecommand \BibitemShut  [1]{\csname bibitem#1\endcsname}%
\bibitem [{\citenamefont {O'Hara}\ \emph {et~al.}(2002)\citenamefont {O'Hara},
  \citenamefont {Hemmer}, \citenamefont {Gehm}, \citenamefont {Granade},\ and\
  \citenamefont {Thomas}}]{O'Hara13122002}%
  \BibitemOpen
  \bibfield  {author} {\bibinfo {author} {\bibfnamefont {K.~M.}\ \bibnamefont
  {O'Hara}}, \bibinfo {author} {\bibfnamefont {S.~L.}\ \bibnamefont {Hemmer}},
  \bibinfo {author} {\bibfnamefont {M.~E.}\ \bibnamefont {Gehm}}, \bibinfo
  {author} {\bibfnamefont {S.~R.}\ \bibnamefont {Granade}}, \ and\ \bibinfo
  {author} {\bibfnamefont {J.~E.}\ \bibnamefont {Thomas}},\ }\Doi
  {10.1126/science.1079107} {\bibfield  {journal} {\bibinfo  {journal}
  {Science},\ }\textbf {\bibinfo {volume} {298}},\ \bibinfo {pages} {2179}
  (\bibinfo {year} {2002})}\BibitemShut {NoStop}%
\bibitem [{\citenamefont {Bourdel}\ \emph {et~al.}(2004)\citenamefont
  {Bourdel}, \citenamefont {Khaykovich}, \citenamefont {Cubizolles},
  \citenamefont {Zhang}, \citenamefont {Chevy}, \citenamefont {Teichmann},
  \citenamefont {Tarruell}, \citenamefont {Kokkelmans},\ and\ \citenamefont
  {Salomon}}]{PhysRevLett.93.050401}%
  \BibitemOpen
  \bibfield  {author} {\bibinfo {author} {\bibfnamefont {T.}~\bibnamefont
  {Bourdel}}, \bibinfo {author} {\bibfnamefont {L.}~\bibnamefont {Khaykovich}},
  \bibinfo {author} {\bibfnamefont {J.}~\bibnamefont {Cubizolles}}, \bibinfo
  {author} {\bibfnamefont {J.}~\bibnamefont {Zhang}}, \bibinfo {author}
  {\bibfnamefont {F.}~\bibnamefont {Chevy}}, \bibinfo {author} {\bibfnamefont
  {M.}~\bibnamefont {Teichmann}}, \bibinfo {author} {\bibfnamefont
  {L.}~\bibnamefont {Tarruell}}, \bibinfo {author} {\bibfnamefont {S.~J. J.
  M.~F.}\ \bibnamefont {Kokkelmans}}, \ and\ \bibinfo {author} {\bibfnamefont
  {C.}~\bibnamefont {Salomon}},\ }\Doi {10.1103/PhysRevLett.93.050401}
  {\bibfield  {journal} {\bibinfo  {journal} {Phys. Rev. Lett.},\ }\textbf
  {\bibinfo {volume} {93}},\ \bibinfo {pages} {050401} (\bibinfo {year}
  {2004})}\BibitemShut {NoStop}%
\bibitem [{\citenamefont {{Regal}}\ \emph {et~al.}(2003)\citenamefont
  {{Regal}}, \citenamefont {{Ticknor}}, \citenamefont {{Bohn}},\ and\
  \citenamefont {{Jin}}}]{2003Natur.424...47R}%
  \BibitemOpen
  \bibfield  {author} {\bibinfo {author} {\bibfnamefont {C.~A.}\ \bibnamefont
  {{Regal}}}, \bibinfo {author} {\bibfnamefont {C.}~\bibnamefont {{Ticknor}}},
  \bibinfo {author} {\bibfnamefont {J.~L.}\ \bibnamefont {{Bohn}}}, \ and\
  \bibinfo {author} {\bibfnamefont {D.~S.}\ \bibnamefont {{Jin}}},\ }\Doi
  {10.1038/nature01738} {\bibfield  {journal} {\bibinfo  {journal} {\nat},\
  }\textbf {\bibinfo {volume} {424}},\ \bibinfo {pages} {47} (\bibinfo {year}
  {2003})},\ \Eprint {http://arxiv.org/abs/arXiv:cond-mat/0305028}
  {arXiv:cond-mat/0305028} \BibitemShut {NoStop}%
\bibitem [{\citenamefont {Strecker}\ \emph {et~al.}(2003)\citenamefont
  {Strecker}, \citenamefont {Partridge},\ and\ \citenamefont
  {Hulet}}]{PhysRevLett.91.080406}%
  \BibitemOpen
  \bibfield  {author} {\bibinfo {author} {\bibfnamefont {K.~E.}\ \bibnamefont
  {Strecker}}, \bibinfo {author} {\bibfnamefont {G.~B.}\ \bibnamefont
  {Partridge}}, \ and\ \bibinfo {author} {\bibfnamefont {R.~G.}\ \bibnamefont
  {Hulet}},\ }\Doi {10.1103/PhysRevLett.91.080406} {\bibfield  {journal}
  {\bibinfo  {journal} {Phys. Rev. Lett.},\ }\textbf {\bibinfo {volume} {91}},\
  \bibinfo {pages} {080406} (\bibinfo {year} {2003})}\BibitemShut {NoStop}%
\bibitem [{\citenamefont {Dieckmann}\ \emph {et~al.}(2002)\citenamefont
  {Dieckmann}, \citenamefont {Stan}, \citenamefont {Gupta}, \citenamefont
  {Hadzibabic}, \citenamefont {Schunck},\ and\ \citenamefont
  {Ketterle}}]{PhysRevLett.89.203201}%
  \BibitemOpen
  \bibfield  {author} {\bibinfo {author} {\bibfnamefont {K.}~\bibnamefont
  {Dieckmann}}, \bibinfo {author} {\bibfnamefont {C.~A.}\ \bibnamefont {Stan}},
  \bibinfo {author} {\bibfnamefont {S.}~\bibnamefont {Gupta}}, \bibinfo
  {author} {\bibfnamefont {Z.}~\bibnamefont {Hadzibabic}}, \bibinfo {author}
  {\bibfnamefont {C.~H.}\ \bibnamefont {Schunck}}, \ and\ \bibinfo {author}
  {\bibfnamefont {W.}~\bibnamefont {Ketterle}},\ }\Doi
  {10.1103/PhysRevLett.89.203201} {\bibfield  {journal} {\bibinfo  {journal}
  {Phys. Rev. Lett.},\ }\textbf {\bibinfo {volume} {89}},\ \bibinfo {pages}
  {203201} (\bibinfo {year} {2002})}\BibitemShut {NoStop}%
\bibitem [{\citenamefont {Giorgini}\ \emph {et~al.}(2008)\citenamefont
  {Giorgini}, \citenamefont {Pitaevskii},\ and\ \citenamefont
  {Stringari}}]{RevModPhys.80.1215}%
  \BibitemOpen
  \bibfield  {author} {\bibinfo {author} {\bibfnamefont {S.}~\bibnamefont
  {Giorgini}}, \bibinfo {author} {\bibfnamefont {L.~P.}\ \bibnamefont
  {Pitaevskii}}, \ and\ \bibinfo {author} {\bibfnamefont {S.}~\bibnamefont
  {Stringari}},\ }\Doi {10.1103/RevModPhys.80.1215} {\bibfield  {journal}
  {\bibinfo  {journal} {Rev. Mod. Phys.},\ }\textbf {\bibinfo {volume} {80}},\
  \bibinfo {pages} {1215} (\bibinfo {year} {2008})}\BibitemShut {NoStop}%
\bibitem [{\citenamefont {Nishida}\ and\ \citenamefont
  {Son}(2007)}]{PhysRevD.76.086004}%
  \BibitemOpen
  \bibfield  {author} {\bibinfo {author} {\bibfnamefont {Y.}~\bibnamefont
  {Nishida}}\ and\ \bibinfo {author} {\bibfnamefont {D.~T.}\ \bibnamefont
  {Son}},\ }\Doi {10.1103/PhysRevD.76.086004} {\bibfield  {journal} {\bibinfo
  {journal} {Phys. Rev. D},\ }\textbf {\bibinfo {volume} {76}},\ \bibinfo
  {pages} {086004} (\bibinfo {year} {2007})}\BibitemShut {NoStop}%
\bibitem [{\citenamefont {Baker}(2000)}]{baker2000mbx}%
  \BibitemOpen
  \bibfield  {author} {\bibinfo {author} {\bibfnamefont {G.}~\bibnamefont
  {Baker}},\ }in\ \href@noop {} {\emph {\bibinfo {booktitle} {Recent progress
  in many-body theories: the proceedings of the 10th international conference,
  Seattle, USA, September 10-15, 1999}}},\ Vol.~\bibinfo {volume} {3}\
  (\bibinfo {organization} {World Scientific Pub Co Inc},\ \bibinfo {year}
  {2000})\ p.~\bibinfo {pages} {15}\BibitemShut {NoStop}%
\bibitem [{\citenamefont {Papenbrock}(2005)}]{PhysRevA.72.041603}%
  \BibitemOpen
  \bibfield  {author} {\bibinfo {author} {\bibfnamefont {T.}~\bibnamefont
  {Papenbrock}},\ }\Doi {10.1103/PhysRevA.72.041603} {\bibfield  {journal}
  {\bibinfo  {journal} {Phys. Rev. A},\ }\textbf {\bibinfo {volume} {72}},\
  \bibinfo {pages} {041603} (\bibinfo {year} {2005})}\BibitemShut {NoStop}%
\bibitem [{\citenamefont {Son}\ and\ \citenamefont
  {Wingate}(2006)}]{Son:2005rv}%
  \BibitemOpen
  \bibfield  {author} {\bibinfo {author} {\bibfnamefont {D.}~\bibnamefont
  {Son}}\ and\ \bibinfo {author} {\bibfnamefont {M.}~\bibnamefont {Wingate}},\
  }\Doi {10.1016/j.aop.2005.11.001} {\bibfield  {journal} {\bibinfo  {journal}
  {Annals Phys.},\ }\textbf {\bibinfo {volume} {321}},\ \bibinfo {pages} {197}
  (\bibinfo {year} {2006})},\ \Eprint {http://arxiv.org/abs/cond-mat/0509786}
  {arXiv:cond-mat/0509786 [cond-mat]} \BibitemShut {NoStop}%
\bibitem [{Note1()}]{Note1}%
  \BibitemOpen
  \bibinfo {note} {See, e.g., \cite {Endres:2012cw} for a extensive summary of
  experimental, theoretical and numerical estimates for $\xi
  _{3d}$.}\BibitemShut {Stop}%
\bibitem [{\citenamefont {Carlson}\ \emph {et~al.}(2011)\citenamefont
  {Carlson}, \citenamefont {Gandolfi}, \citenamefont {Schmidt},\ and\
  \citenamefont {Zhang}}]{PhysRevA.84.061602}%
  \BibitemOpen
  \bibfield  {author} {\bibinfo {author} {\bibfnamefont {J.}~\bibnamefont
  {Carlson}}, \bibinfo {author} {\bibfnamefont {S.}~\bibnamefont {Gandolfi}},
  \bibinfo {author} {\bibfnamefont {K.~E.}\ \bibnamefont {Schmidt}}, \ and\
  \bibinfo {author} {\bibfnamefont {S.}~\bibnamefont {Zhang}},\ }\Doi
  {10.1103/PhysRevA.84.061602} {\bibfield  {journal} {\bibinfo  {journal}
  {Phys. Rev. A},\ }\textbf {\bibinfo {volume} {84}},\ \bibinfo {pages}
  {061602} (\bibinfo {year} {2011})}\BibitemShut {NoStop}%
\bibitem [{\citenamefont {Ku}\ \emph {et~al.}(2012)\citenamefont {Ku},
  \citenamefont {Sommer}, \citenamefont {Cheuk},\ and\ \citenamefont
  {Zwierlein}}]{Ku03022012}%
  \BibitemOpen
  \bibfield  {author} {\bibinfo {author} {\bibfnamefont {M.~J.~H.}\
  \bibnamefont {Ku}}, \bibinfo {author} {\bibfnamefont {A.~T.}\ \bibnamefont
  {Sommer}}, \bibinfo {author} {\bibfnamefont {L.~W.}\ \bibnamefont {Cheuk}}, \
  and\ \bibinfo {author} {\bibfnamefont {M.~W.}\ \bibnamefont {Zwierlein}},\
  }\Doi {10.1126/science.1214987} {\bibfield  {journal} {\bibinfo  {journal}
  {Science},\ }\textbf {\bibinfo {volume} {335}},\ \bibinfo {pages} {563}
  (\bibinfo {year} {2012})}\BibitemShut {NoStop}%
\bibitem [{\citenamefont {Nishida}\ and\ \citenamefont
  {Tan}(2008)}]{PhysRevLett.101.170401}%
  \BibitemOpen
  \bibfield  {author} {\bibinfo {author} {\bibfnamefont {Y.}~\bibnamefont
  {Nishida}}\ and\ \bibinfo {author} {\bibfnamefont {S.}~\bibnamefont {Tan}},\
  }\Doi {10.1103/PhysRevLett.101.170401} {\bibfield  {journal} {\bibinfo
  {journal} {Phys. Rev. Lett.},\ }\textbf {\bibinfo {volume} {101}},\ \bibinfo
  {pages} {170401} (\bibinfo {year} {2008})}\BibitemShut {NoStop}%
\bibitem [{\citenamefont {Nishida}\ and\ \citenamefont
  {Son}(2010)}]{Nishida:2009pg}%
  \BibitemOpen
  \bibfield  {author} {\bibinfo {author} {\bibfnamefont {Y.}~\bibnamefont
  {Nishida}}\ and\ \bibinfo {author} {\bibfnamefont {D.~T.}\ \bibnamefont
  {Son}},\ }\Doi {10.1103/PhysRevA.82.043606} {\bibfield  {journal} {\bibinfo
  {journal} {Phys.Rev.},\ }\textbf {\bibinfo {volume} {A82}},\ \bibinfo {pages}
  {043606} (\bibinfo {year} {2010})},\ \Eprint {http://arxiv.org/abs/0908.2159}
  {arXiv:0908.2159 [cond-mat.quant-gas]} \BibitemShut {NoStop}%
\bibitem [{\citenamefont {{Tan}}(2008){\natexlab{a}}}]{2008AnPhy.323.2952T}%
  \BibitemOpen
  \bibfield  {author} {\bibinfo {author} {\bibfnamefont {S.}~\bibnamefont
  {{Tan}}},\ }\Doi {10.1016/j.aop.2008.03.004} {\bibfield  {journal} {\bibinfo
  {journal} {Annals of Physics},\ }\textbf {\bibinfo {volume} {323}},\ \bibinfo
  {pages} {2952} (\bibinfo {year} {2008}{\natexlab{a}})},\ \Eprint
  {http://arxiv.org/abs/cond-mat/0505200} {cond-mat/0505200} \BibitemShut
  {NoStop}%
\bibitem [{\citenamefont {{Tan}}(2008){\natexlab{b}}}]{2008AnPhy.323.2971T}%
  \BibitemOpen
  \bibfield  {author} {\bibinfo {author} {\bibfnamefont {S.}~\bibnamefont
  {{Tan}}},\ }\Doi {10.1016/j.aop.2008.03.005} {\bibfield  {journal} {\bibinfo
  {journal} {Annals of Physics},\ }\textbf {\bibinfo {volume} {323}},\ \bibinfo
  {pages} {2971} (\bibinfo {year} {2008}{\natexlab{b}})},\ \Eprint
  {http://arxiv.org/abs/cond-mat/0508320} {cond-mat/0508320} \BibitemShut
  {NoStop}%
\bibitem [{\citenamefont {{Tan}}(2008){\natexlab{c}}}]{2008AnPhy.323.2987T}%
  \BibitemOpen
  \bibfield  {author} {\bibinfo {author} {\bibfnamefont {S.}~\bibnamefont
  {{Tan}}},\ }\Doi {10.1016/j.aop.2008.03.003} {\bibfield  {journal} {\bibinfo
  {journal} {Annals of Physics},\ }\textbf {\bibinfo {volume} {323}},\ \bibinfo
  {pages} {2987} (\bibinfo {year} {2008}{\natexlab{c}})},\ \Eprint
  {http://arxiv.org/abs/arXiv:0803.0841} {arXiv:arXiv:0803.0841
  [cond-mat.stat-mech]} \BibitemShut {NoStop}%
\bibitem [{\citenamefont {Braaten}\ and\ \citenamefont
  {Platter}(2008)}]{PhysRevLett.100.205301}%
  \BibitemOpen
  \bibfield  {author} {\bibinfo {author} {\bibfnamefont {E.}~\bibnamefont
  {Braaten}}\ and\ \bibinfo {author} {\bibfnamefont {L.}~\bibnamefont
  {Platter}},\ }\Doi {10.1103/PhysRevLett.100.205301} {\bibfield  {journal}
  {\bibinfo  {journal} {Phys. Rev. Lett.},\ }\textbf {\bibinfo {volume}
  {100}},\ \bibinfo {pages} {205301} (\bibinfo {year} {2008})}\BibitemShut
  {NoStop}%
\bibitem [{\citenamefont {{Braaten}}\ \emph {et~al.}(2008)\citenamefont
  {{Braaten}}, \citenamefont {{Kang}},\ and\ \citenamefont
  {{Platter}}}]{2008PhRvA..78e3606B}%
  \BibitemOpen
  \bibfield  {author} {\bibinfo {author} {\bibfnamefont {E.}~\bibnamefont
  {{Braaten}}}, \bibinfo {author} {\bibfnamefont {D.}~\bibnamefont {{Kang}}}, \
  and\ \bibinfo {author} {\bibfnamefont {L.}~\bibnamefont {{Platter}}},\ }\Doi
  {10.1103/PhysRevA.78.053606} {\bibfield  {journal} {\bibinfo  {journal}
  {\pra},\ }\textbf {\bibinfo {volume} {78}},\ \bibinfo {eid} {053606}
  (\bibinfo {year} {2008})},\ \Eprint {http://arxiv.org/abs/0806.2277}
  {arXiv:0806.2277 [cond-mat.other]} \BibitemShut {NoStop}%
\bibitem [{\citenamefont {Chen}\ and\ \citenamefont
  {Kaplan}(2004)}]{PhysRevLett.92.257002}%
  \BibitemOpen
  \bibfield  {author} {\bibinfo {author} {\bibfnamefont {J.-W.}\ \bibnamefont
  {Chen}}\ and\ \bibinfo {author} {\bibfnamefont {D.~B.}\ \bibnamefont
  {Kaplan}},\ }\Doi {10.1103/PhysRevLett.92.257002} {\bibfield  {journal}
  {\bibinfo  {journal} {Phys. Rev. Lett.},\ }\textbf {\bibinfo {volume} {92}},\
  \bibinfo {pages} {257002} (\bibinfo {year} {2004})}\BibitemShut {NoStop}%
\bibitem [{\citenamefont {Symanzik}(1983){\natexlab{a}}}]{Symanzik1983187}%
  \BibitemOpen
  \bibfield  {author} {\bibinfo {author} {\bibfnamefont {K.}~\bibnamefont
  {Symanzik}},\ }\Doi {10.1016/0550-3213(83)90468-6} {\bibfield  {journal}
  {\bibinfo  {journal} {Nuclear Physics B},\ }\textbf {\bibinfo {volume}
  {226}},\ \bibinfo {pages} {187 } (\bibinfo {year} {1983}{\natexlab{a}})},\
  ISSN \bibinfo {issn} {0550-3213}\BibitemShut {NoStop}%
\bibitem [{\citenamefont {Symanzik}(1983){\natexlab{b}}}]{Symanzik1983205}%
  \BibitemOpen
  \bibfield  {author} {\bibinfo {author} {\bibfnamefont {K.}~\bibnamefont
  {Symanzik}},\ }\Doi {10.1016/0550-3213(83)90469-8} {\bibfield  {journal}
  {\bibinfo  {journal} {Nuclear Physics B},\ }\textbf {\bibinfo {volume}
  {226}},\ \bibinfo {pages} {205 } (\bibinfo {year} {1983}{\natexlab{b}})},\
  ISSN \bibinfo {issn} {0550-3213}\BibitemShut {NoStop}%
\bibitem [{\citenamefont {Endres}\ \emph {et~al.}(2012)\citenamefont {Endres},
  \citenamefont {Kaplan}, \citenamefont {Lee},\ and\ \citenamefont
  {Nicholson}}]{Endres:2012cw}%
  \BibitemOpen
  \bibfield  {author} {\bibinfo {author} {\bibfnamefont {M.~G.}\ \bibnamefont
  {Endres}}, \bibinfo {author} {\bibfnamefont {D.~B.}\ \bibnamefont {Kaplan}},
  \bibinfo {author} {\bibfnamefont {J.-W.}\ \bibnamefont {Lee}}, \ and\
  \bibinfo {author} {\bibfnamefont {A.~N.}\ \bibnamefont {Nicholson}},\
  }\href@noop {} { (\bibinfo {year} {2012})},\ \Eprint
  {http://arxiv.org/abs/1203.3169} {arXiv:1203.3169 [hep-lat]} \BibitemShut
  {NoStop}%
\bibitem [{Note2()}]{Note2}%
  \BibitemOpen
  \bibinfo {note} {Note that $\protect \mathcal O_j$ are implicit functions of
  the fixed dimensionless quantity $\protect \mathaccentV {hat}05Em
  $.}\BibitemShut {Stop}%
\bibitem [{Note3()}]{Note3}%
  \BibitemOpen
  \bibinfo {note} {Here, I assume the canonical partition function has been
  normalized such that $F(0) = 0$}\BibitemShut {NoStop}%
\bibitem [{\citenamefont {{Endres}}(2012)}]{2012PhRvA..85f3624E}%
  \BibitemOpen
  \bibfield  {author} {\bibinfo {author} {\bibfnamefont {M.~G.}\ \bibnamefont
  {{Endres}}},\ }\Doi {10.1103/PhysRevA.85.063624} {\bibfield  {journal}
  {\bibinfo  {journal} {\pra},\ }\textbf {\bibinfo {volume} {85}},\ \bibinfo
  {eid} {063624} (\bibinfo {year} {2012})},\ \Eprint
  {http://arxiv.org/abs/1204.6182} {arXiv:1204.6182 [hep-lat]} \BibitemShut
  {NoStop}%
\end{thebibliography}%

\end{document}